\documentclass[preprint,aps,floats,superscriptaddress]{revtex4}
\usepackage{amsmath}
\usepackage{graphicx}
\usepackage{epsfig}

\def\figx#1#2{\includegraphics[width=#1]{#2}}

\newcommand{\si}{\sigma}
\newcommand{\dg}{^{\dagger}}

\newcommand{\la}{\langle}
\newcommand{\ra}{\rangle}

\newcommand{\rarrow}{\rightarrow }

\newlength{\bxwidth}
\newcommand\frm[1]{\epsfig{file=#1,width=\bxwidth}}
\newlength{\fight}
\newcommand{\fg}[3]
{\begin{figure}[here]
\[ 
\figx{\fight}{#1}
\] 
\vspace*{-4mm}
\caption{\label{#2}
\small #3}
\end{figure} }

\begin{document}

\title{How do Fermi liquids get heavy and die?}

\author{P. Coleman}
\affiliation{Center for Materials Theory,
Department of Physics and Astronomy, Rutgers University, Piscataway, NJ
08854, USA}

\author{C. P{\'e}pin}
\affiliation{SPhT, L'Orme des Merisiers, CEA-Saclay, 91191 Gif-sur-Yvette, France.}  

\author{Qimiao Si}
\affiliation{Department of Physics and Astronomy, Rice University,
Houston, TX 77251, USA}  

\author{R. Ramazashvili}
\affiliation{Department of Physics, University of
Illinois, Urbana, IL 61801, U.S.A.}  

\begin{abstract}
We discuss non-Fermi liquid and quantum
critical behavior in heavy fermion materials, focussing 
on the mechanism by which the
electron mass appears to diverge at the quantum critical point.
We ask whether the basic mechanism for the transformation involves
electron diffraction off a quantum critical spin density wave,
or whether a break-down in the composite nature of the heavy electron
takes place at the quantum critical point. 
We show that the Hall constant
changes continously in the first scenario, but may
``jump'' discontinuously at a quantum critical point where the
composite character of the electron quasiparticles
changes.
\end{abstract}

\maketitle

%%  {\sl
%%  Outline:
%%  \begin{itemize}
%%  
%%  \item Brief history of mass renormalization.  Spin-fluctuations.
%%  Kondo effect and Heavy fermions.  QCP $\rarrow $  break-down of Fermi liquid
%%  behavior. 
%%  
%%  \item Millis-Hertz vs fundamental breakdown. How to frame the question
%%  in language that can be addressed by definite experiments and
%%  concrete calculations? 
%%  
%%  \item The two pictures Hot lines versus entire Fermi-surface.
%%  Local versus global divergence of mass. Folding of the Fermi surface
%%  versus complete change in topology. 
%%  
%%  \item Does the linear specific heat diverge? 
%%  
%%  \item Is there scaling? Upper critical dimensions.
%%  
%%  \item Does the Fermi surface geometry change? - Hall constant as a probe.
%%  
%%  \item How does the Hall constant change in (i) a spin density folding 
%%  or (ii) a suddent topological transformation of the Fermi surface? 
%%  
%%  
%%  \item Speculation: if not spin fluctuations- low-lying
%%  excitations are those associated with fluctuations in the topology
%%  of the Fermi surface.   What is the nature of these strong fluctuations?
%%  Often assumed to be bosonic collective modes- but difficult
%%  to avoid a return to Fermi liquid and 
%%  Millis Hertz in 3D.  These are modes which most-likely change
%%  the Luttinger sum rule are more likely to carry the character
%%  of fermionic resonances. 
%%  
%%   
%%  \end{itemize}
%%  }
%%  \newpage
%%  
%intro

\subsubsection{Introduction: Mass divergence and the break-down of the
quasiparticle concept. }

A key element in the Landau Fermi liquid theory\cite{landau,early,early2} is the  idea
of quasiparticles: excitations of the Fermi
sea that carry the original charge  and spin quantum numbers of the
non-interacting particles from which they are derived, but whose mass $m^*$ is
renormalized by interactions. 
What is the fate of quasiparticles 
when interactions become so large
that the ground state is no longer adiabatically connected
to a non-interacting system? 
It is known 
that the quasi-particle mass
diverges in the approach
to a zero temperature ferromagnetic instability\cite{brinkmanrice,volhardt,paladium,moriya}.
Recent measurements on
three dimensional heavy fermion compounds suggest\cite{steglich3,aoki,hilbert1,hilbert2,schroeder} that the quasiparticle
mass also diverges in the approach to an antiferromagnetic
quantum critical point. A central property of the Landau
quasiparticle, is the existence of a finite overlap ``$Z$'', or ``wavefunction
renormalization''  between
a single quasiparticle state, denoted by  $\vert \hbox{qp}^{-}\ra$ and the state formed by
adding a single electron  to the ground-state, denoted by $\vert e^{-
}\ra = c\dg _{{\bf k }\sigma}\vert 0\ra$, 
\begin{equation}\label{}
Z= \vert \la e^{- }\vert \hbox{qp}^{-}\ra \vert ^{2}.
\end{equation}
This quantity is closely related
to the ratio $m/m^{*}$ of the electron to quasiparticle mass,  $Z\sim
m/m^{*}$ and if the quasiparticle mass diverges,  the overlap
between the quasiparticle and the electron state from which it is
derived is driven to zero, signalling a break-down in the
quasiparticle concept.  Thus 
the divergence  of the electron mass at an antiferromagnetic
quantum critical point (QCP)
has important consequences, for it indicates that 
antiferromagnetism causes a break-down in the Fermi liquid concept\cite{varma2001}.

In the late seventies, unprecedented 
mass renormalization
was discovered in heavy fermion compounds.
In  these  materials,
quasiparticle masses of  order $100$, but sometimes in  excess of $1000$ bare
electron  masses  have  been recorded, a significant fraction
of which is thought to derive from their close 
vicinity to an antiferromagnetic
quantum critical point. 
The discovery of the cuprate superconductors in the late eighties 
further sharpened interest in the effects of strong antiferromagnetic
interactions. In these materials, 
unconventional normal state properties 
have led many to believe that a combination of 
strong antiferromagnetic
correlations and low dimensionality may lead to a complete 
break-down 
\begin{figure}[here]
\[ 
\figx{9.5cm}{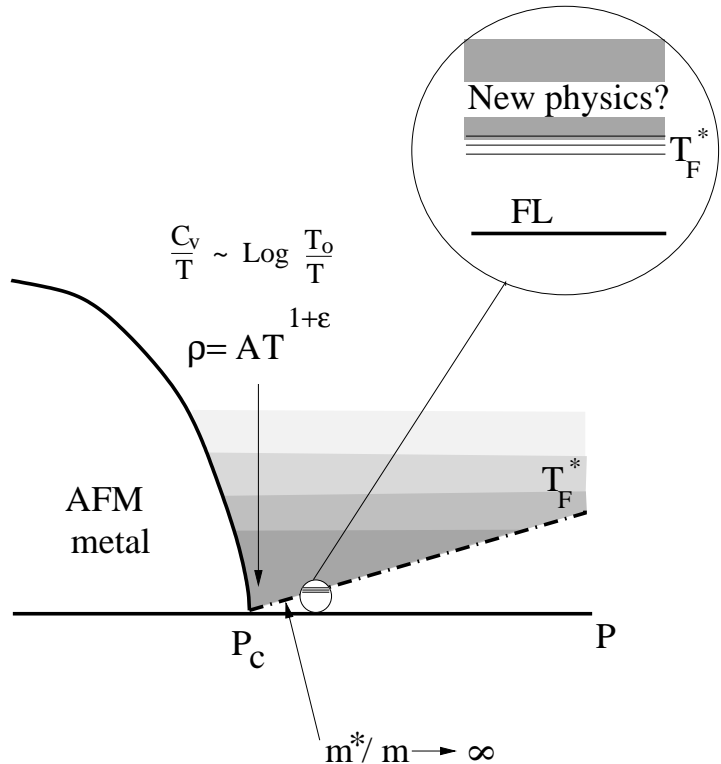}
\] 
\vspace*{-4mm}
\caption{\label{fig1}
\small
Schematic illustration of quantum critical point physics
in heavy fermion metals. In cerium and uranium heavy fermion systems,
pressure drives the system from the antiferromagnet, into the
paramagnetic phase. In ytterbium systems, the situation is reversed,
and pressure induces antiferromagnetism. The inset emphasizes the
point that if the Fermi temperature goes to zero at the quantum
critical point, then a new class of universal excitations is expected
above $T_{F}^{*}$.
}\end{figure} 
\noindent 
of  the electron
quasiparticle.  Against this backdrop,
heavy fermion materials acquire a new significance. 
The appearance of non-Fermi liquid behavior, often  in conjunction
with anisotropic superconductivity near a heavy fermion quantum critical point
tempts us to believe that there may be certain aspects 
of the quantum critical physics that are shared 
between heavy fermion and cuprate materials.\cite{varma1} The clear advantage of 
heavy fermions lies in the ability 
to use pressure
\cite{mathur} or chemical pressure \cite{hilbert1,hilbert2,schroeder}, to tune
them continuously from an unambiguous three dimensional fermi liquid into an 
antiferromagnetic quantum critical point where non-Fermi liquid
behavior is consistently manifested, permitting a first
systematic study of the break-down of the Fermi liquid in the presence
of critical antiferromagnetism (see figure 1).

\subsubsection{Properties of the Heavy Fermion Quantum Critical Point}
\vspace{0.1 truein}
\begin{center}
\centerline{\bf Table. 1. Selected Heavy Fermion compounds with quantum critical points.}
\vskip 0.1 truein
\begin{tabular}{|c||c|c|c|c|}
\hline
Compound & $P_c$/$x_c$ & $\frac{C_v}{T}\rarrow \infty $? & $\rho\sim T^a $ & Reference\\
\hline
$CeNi_2Ge_2$ & $P_c =0$  & $Log \left( \frac{T_o}{T}\right)$$^{(a)}$
& $T^{1.2}$ & \cite{grosche,gegenwart,julian,braithwaite} \\
\hline
$YbRh_2(Si_{1-x}Ge_x)_2$ & $x_c=0.05$ &
$Log \left( \frac{T_o}{T}\right)$
$^{(b)}$ & $ T$ & \cite{steglich} \\
\hline
$Ce(Cu_{1-x}Au_x)_6 $ & $x_c=0.016$ & $Log \left( \frac{T_o}{T}\right)$ & $ T + c $ & \cite{hilbert1,hilbert2,schroeder} \\
\hline
$CeCu_{6-x}Ag_x$ & $x_c=0.09$ & $Log \left( \frac{T_o}{T}\right)$ & $T^{1.1}$ & \cite{heuser}\\
\hline
$U_2Pt_2In$ & $P_c=0$  & $Log \left( \frac{T_o}{T}\right)$
 & $T$ & \cite{devisser}\\
\hline
$U_2Pd_2 In$ & $P_c<0 $ &  ? & $T + c $ & \cite{devisser} \\
\hline
$CePd_2Si_2$ & $ P_c>0$  & ? & $ T^{1.2}$ & \cite{grosche} \\
\hline
$CeIn_3$ & $P_c>0$ & ? & $T^{1.5} $ & \cite{mathur,grosche} \\
\hline
$U_3Ni_3Sn_4$ & $P_c >0$ & no & ? & \cite{devisser2} \\
\hline
$CeCu_2Si_2$ & $Pc=0$ & no & $T^{1.5}$ & \cite{cecu2si2}\\
\hline
\hline
\end{tabular}

\end{center}
\vspace{0.1 truein}
{\small $^{(a)}$ New  data\cite{newgegen} show a stronger divergence at lower
temperatures, and $\gamma \sim A- B \sqrt{T}$ at intermediate temperatures.

\noindent $^{(b)}$ At low temperatures, $\gamma$ diverges more rapidly than 
 $Log \left( \frac{T}{T_o}\right)$\cite{steglich}.
}\\

\vspace{0.3 truein}
There are several heavy fermion systems that
have been tuned to an antiferromagnetic quantum critical point.(See table 1.)
Two stoichiometric heavy fermion systems, $CeNi_{2}Ge_{2}$\cite{grosche},
$U_{2}Pt_{2}In$\cite{devisser} 
lie almost at a quantum critical point at ambient pressure, whilst the
compounds  $CeCu_{6}$\cite{schroeder,heuser}
and $YbRh_{2}Si_{2}$\cite{steglich} can be tuned to a quantum critical
point with a tiny amount of chemical pressure, applied by doping. 
%The paramagnetic system $Ce 
%Cu_{6-x}Au_{x}$\cite{schroeder} can 
%be
%tuned to a quantum critical point at $x=0.1$, using (negative)
%chemical pressure applied by doping with gold. {
%\it 
%$CeCu_{6-x}Ag_x$ 
%shows similar behaviour upon doping with silver, with a quantum critical point at $x=0.09$~\cite{heuser} }. In the case of 
%$YbRh_{2} ( Si_{1-x}Ge_{x})_{2}$, chemical pressure acts to drive
%the system towards the paramagnetic state, and here, five percent 
%($x=0.05$) Germanium 
%is sufficient to  fine tune a  tiny $60mK$ 
%antiferromagnetic transition temperature to zero, 
%without significantly affecting the residual resistivity. 
There is a growing list of 
antiferromagnetic cerium and uranium systems that can be driven
paramagnetic by the application of pressure, including
$CePd_{2}Si_{2}$\cite{mathur}, $U_{2}Pd_{2}In$\cite{devisser},
$CeIn_{3}$\cite{mathur} 
and
$CeRh_{2}Si_{2}$\cite{movshkovich}.  Diamond anvil methods
%\cite{pugh} 
are expected
to substantially add to this list in the near future.

{%\it 
By its very nature, quantum criticality is
hyper-sensitive to disorder, which can never be totally eliminated~\cite{disorder}.
Nevertheless, the quantum critical point in
the idealized limit of a perfectly clean system poses an intellectual
challenge in its own right that must be understood before moving
on to the more complex effects of disorder.
%With this guiding philosophy in mind, we will not discuss those materials 
%far from stoichiometry in which 
%large inhomogenieties are suspected to be present, or those materials
%which have not been explicitly tuned 
%from the antiferromagnetic to paramagnetic phase using 
%pressure or modest amounts ($<10\%$) of chemical doping. 
With this guiding philosophy in mind, we will {\it not} 
discuss those materials far from stoichiometry, in which 
large inhomogeneities are suspected. 
The discussion here will focus on materials that are
explicitly stoichiometric and tuned to the critical point 
by pressure, or those where critical behavior is insensitive 
to whether small doping or pressure was used to tune them 
to criticality. These materials show many common properties: 

}

The key properties of these compounds are:

\begin{itemize}

\item Fermi liquid
behavior  in the paramagnetic phase.\cite{devisser2,flouquet}   One of the classic
features of Fermi liquid behavior, a 
quadratic temperature dependence of the resistivity 
$\rho =\rho_{o}+A T^{2}$ develops at ever lower temperatures, 
in the approach to the 
the quantum phase transition, 
indicating that  Fermi liquid behavior
survives right up to the quantum critical point. 

\item 
Divergent  specific heat coefficients
at the critical point. In many cases, the divergence 
displays a logarithmic temperature dependence, 
\begin{equation}\label{}
\gamma (T) = \frac{C_{v} (T)}{T} = \gamma _{0} \log \left[\frac{T_{o}}{T} \right].
\end{equation}
This suggests that the Fermi temperature renormalizes
to zero and the quasiparticle effective masses diverge 
\begin{equation}\label{}
T_{F}^{*}\rarrow 0, \qquad \frac{m^{*}}{m}\rarrow \infty 
\end{equation}
at the quantum critical point of these three dimensional materials.
Further 
support for this conclusion is provided by the observation that
the quadratic coefficient $A$ of  the resistivity  grows
in the approach to the quantum critical point\cite{devisser}.
As yet however, surprisingly little is known about the variation of the
zero-temperature 
linear specific heat coefficient in the approach to the quantum critical point.

\item Quasi-linear temperature dependence in the resistivity
\begin{equation}\label{}
\rho \propto T^{1+\epsilon},
\end{equation}
with
$\epsilon$ in the range $ 0-0.6$. 
The most impressive results
to date have been observed in $YbRh_{2}Si_{2}$, where a linear
resistivity over three decades develops at the quantum critical point\cite{steglich}.

\item Non-Curie spin susceptibilities
\begin{equation}\label{}
\chi ^{-1} (T)= \chi _{0}^{-1}
+ c T^{a} 
\end{equation}
with $a<1$ observed in critical 
$Ce Cu_{6-x}Au_{x}$ (x=0.1), $YbRh_{2} (Si_{1-x}Ge_{x})_{2}$ (x=0.05)
and $CeNi_{2}Ge_{2}$. 
In critical $Ce Cu_{6-x}Au_{x}$, the differential magnetic
susceptibility
$dM/dB$
also exhibits $B/T$ scaling in the form 
\begin{equation}\label{}
(dM/dB)^{-1} = \chi _{0}^{-1}
+ c T^{a} g[B/T], 
\end{equation}
and extensive inelastic neutron measurements
show that 
the dynamical spin susceptiblity 
exhibits $E/T$ scaling\cite{varma1989,aronson} throughout the Brillouin zone, parameterized in the form
\begin{equation}\label{lab1}
\chi^{-1} ({\bf q},\omega ) =  T^{a}f (E/T)+\chi _{0}^{-1} ({\bf q})
\end{equation}
where $a\approx 0.75$ and $F[x]\propto (1-ix)^{a}$.  Scaling behavior
with a single exponent in 
the momentum-independent
component of the dynamical spin susceptibility 
suggests an emergence of {\sl local} magnetic moments  which 
are {\sl critically correlated in time} at the quantum critical point. 
$E/T$  and $B/T$ scaling,  with unusual
exponents, represent important 
signatures of universal critical fluctuations, as we now discuss.

\end{itemize} 

\subsubsection{Is there Universality at a Heavy Fermion QCP?}

Usually, the physics of a metal above its
Fermi temperature depends on the 
detailed chemistry and
band-structure of the material: it is non-universal. 
A divergence in the linear specific heat of heavy fermion systems at
the quantum critical point offers the possibility of a very 
different state of affairs, 
for if the renormalized Fermi temperature $T^{*}_{F}
(P)$ can be tuned to become arbitrarily small compared with the 
characteristic scales of the material, 
we expect that 
the ``high energy'' physics
{\sl above} the Fermi temperature $T^{*}_{F}$ is itself, 
\underline{universal}. This is 
an unusual  situation, more akin to that 
in particle
physics.
This  potential
for universal fixed point physics above the Fermi temperature
is of particular interest, for we expect
that like a Fermi liquid, it should involve
a robust set of universal excitations, or quasiparticles, describing the
emergence of magnetism, whose
interactions and energies only depend on the symmetry of
the crystal plus a small set of relevant parameters.

%The significance of universality in the context of
%quantum criticality deserves special digression. 
In classical statistical mechanics, universality
manifests itself through the appearance of scaling laws
and critical exponents that are so robust
to details of the underlying physics, that 
they re-occur in such diverse contexts as 
the critical 
point of water and the Curie point of a ferromagnet. 
Dimensionality 
plays a central role in this universality. 
Provided that the dimensionality of the classical critical point
lies below its ``upper critical dimension'', then 
the thermodynamics and correlation functions near 
the critical point are dominated by a 
single length scale. The emergence of a single length in the correlation functions and
thermodynamics is called ``hyperscaling''. At the critical point,
the Fourier transformed correlation function takes the form
%%%CHANGE q to q-Q and DEFINE Q
\begin{equation}\label{}
S ({\bf q})\propto \frac{1}{|{\bf q}-{\bf Q}|^{2-\eta}} \ ,
\end{equation}
where  ${\bf Q}$ is the ordering wavevector.
%where $q$ is the momentum. 
Suppose the system has finite spatial extent $L$ in one or more directions. 
Below the upper-critical dimension,
$L$ is the only spatial scale in the problem, and the correlation
function 
now develops a finite correlation length $\xi\propto L$, so that
\begin{equation}\label{}
S (q)= \frac{1}{q^{2-\eta}} F (qL).
\end{equation}
where $F(x)$ is a universal function of dimensionless parameters. 
The short-distance physics does not affect the finite correlation length
induced by the finite size.

By contrast, if the system is above the  upper critical dimension, 
the quantum critical point is
no longer dominated by a single length scale and  ``naive''
scaling laws involve corrections associated with the
short-range interaction between  the critical modes. 
\cite{zinnjustin}. For example, the critical theory governing the
liquid gas critical point, the so called $\phi^4$ theory has an upper
critical dimension $d_{u}=4$. Above four dimensions, the 
short-range interactions, denoted by a parameter   $U$ 
affect the correlation length
when the system is near the critical point, so that now
the correlation function above the upper critical dimension takes the form
\begin{equation}\label{}
S (q)= \frac{1}{q^{2}} F (q\xi).
\end{equation}
where the correlation length is now {\it more}  than $L$, determined by a
function of the form
\begin{equation}\label{}
\xi^{-1}= L^{-1}G(U,L)
\end{equation}
where $G(x)$ is a dimensionless function determined by the Gaussian
fluctuations about the mean-field theory.
Indeed, above $d=4$, the scaling dimensions of $U$ are $[U]=L^{d-4}$,
so that $U$ and $L$ must enter in the combination $U/L^{d-4}$. Detailed
calculations show that $G\propto U^{\frac{1}{2}}$, so that 
\begin{equation}\label{}
\xi^{-1}\propto  L^{-1}\left(\frac{U}{L^{d-4}} \right)^{\frac{1}{2}}
\end{equation}
above the upper critical dimension $d=4$.

Universality in the context of quantum criticality 
implies the extension  of these same
principles to the quantum fluctuations that develop at a second-order 
instability in the ground-state. 
Quantum critical behavior implies a divergence in both 
the long distance and long-time correlations in the material.
In quantum statistical mechanics, 
temperature provides a natural
cutoff timescale
\begin{equation}\label{}
\tau_{T} = \frac{\hbar }{k_{B}T}
\end{equation}
beyond which coherent quantum processes  are dephased by thermal
fluctuations. We are thus dealing with a problem of finite size
scaling in the time direction\cite{zinnjustin}.
If a quantum critical system exhibits hyperscaling, then $\tau_T$ must
set the temporal correlation length $\tau$, i.e $\tau \propto \tau
_{T}$, so that at the quantum critical point the correlation 
functions in the frequency domain take 
the form
\begin{equation}\label{}
F (\omega,T)= \frac{1}{\omega^{\alpha }} f (\omega \tau )= 
\frac{1}{\omega^{\alpha }} f (\hbar \omega /k_{B}T).
\end{equation}
Since dynamic response functions at energy $E=\hbar \omega$
are directly proportional to correlation functions at frequency
$\omega$ via the fluctuation dissipation theorem, 
it follows that 
dynamic response functions 
at a quantum critical point below its upper critical dimension
are expected to obey $E/T $
scaling\cite{sachdev}
\begin{equation}\label{}
F (E,T)= \frac{1}{E^{\alpha }} f (E/k_{B}T)
\end{equation}

Above the upper critical dimension, naive scaling no longer
applies. In this case the strength of the short-range interactions between
the  critical modes play the role of ``dangerous irrelevant
variables''
which affect the correlation time\cite{sachdevbook}. 
In a quantum  $\phi ^{4}$ or ``Hertz Millis'' field theory,\cite{hertz,millis}, the temporal correlation time above the upper critical dimension
takes the form
\begin{equation}\label{}
\tau^{-1} =\tau _{T}^{-1} R(T,U)
\end{equation}
where $R(T,U)$ is a dimensionless universal function.
Near a quantum critical point, the dynamical correlation time $\tau $
scales with the spatial correlation length $\xi$ through a dynamical exponent $z$ such that
\begin{equation}\label{} \tau ^{-1}\propto \xi^{-z} \ .
\end{equation}
For a ``Hertz Millis'' theory of a critical spin density wave\cite{hertz,millis}, 
$z=2$ and  $T$ and $U$ enter into 
$R$ as the dimensionless combination $U T^{(d+z-4)/z} = U
T^{\frac{1}{2}}$. Furthermore,  $R(X)\propto X$, thus
$\tau ^{-1}\sim UT^{3/2}$ in three spatial
dimensions, corresponding  to $E/T^{3/2}$ scaling\cite{sachdevbook}. 
\begin{equation}\label{}
F (E,T)= \frac{1}{E^{\alpha }} f (E\tau )= 
\frac{1}{E^{\alpha }} f (E/T^{3/2}).
\end{equation}
For a generic effective Lagrangian $\omega/T$ scaling will not 
occur above the upper critical dimension\cite{raymond}. 
Thus the observation of $E/T$ scaling in the dynamical spin
susceptibility indicates that the underlying physical
theory lies beneath its upper critical dimension.
What is this 
universal ``non-Fermi liquid'' physics,  and what is 
the mechanism by which the mass of the heavy electrons diverges
in the approach to the anti-ferromagnetic instability? 

The existence of a Fermi liquid 
either side of the antiferromagnetic quantum critical point in heavy fermion
materials affords a unique perspective on the above
question, for it tells us that the universal Lagrangian governing the
quantum criticality must find expression in terms of fields
that describe the quasiparticles in the Fermi liquid. 
If we write the low-energy physics of the Fermi liquid in terms of a
Lagrangian, we expect to be able to divide it into three terms
\begin{equation}\label{lag}
L = L_{F} + L_{F-M} + L_{M}.
\end{equation}
where $L_{F}$ describes the free energy of the 
paramagnetic Fermi liquid,  far from the magnetic
instability: this term would involve the short-range interactions between
the quasiparticles and the band-structure. The last term,
$L_{M}$ describes the magnetic excitations that 
emerge above the energy scale $T_{F}^{*} (P)$ and $L_{F-M}$ describes
the way that the quasiparticles couple to and decay into these
magnetic modes. We may then ask:
\begin{enumerate}

\item what is the 
nature of the quantum fields that carry the magnetism, whose activation
at energy scales above $T_{F}^{*} (P)$ is described by 
the Lagrangian $L_{M}$, and 

\item what are the interaction terms $L_{F-M}$  that couple the
low-energy quasiparticles to these universal, high energy
excitations? 

\end{enumerate}

\subsubsection{Spin Density Waves versus Composite Quasiparticles}

We shall now contrast two competing answers 
to this question, one in which 
non-Fermi liquid  behavior derives from Bragg diffraction of the electrons
off a critical spin density wave, the other in which 
the bound-state structure of the composite heavy fermions breaks down at the
QCP. 

If we suppose first 
that the QCP is 
a spin-density wave instability\cite{overhauser}
of the Fermi surface, then  non-Fermi liquid behavior results from
the Bragg scattering of electrons off a critical spin density wave.
In this ``weak-coupling'' approach $L_{F-M}$  is a classical coupling between the 
modes of a spin density wave and the 
Fermi liquid
\begin{eqnarray}\label{weak}
L_{F-M}^{(1)}&=& g \sum_{{\bf k},{\bf q} } 
\vec{\sigma}_{-{\bf q}}\cdot \vec{M}_{{\bf q}}
\end{eqnarray}
where $\vec{\sigma} _{-{\bf q} }=\sum_{{\bf k}} c\dg _{{\bf k}-{\bf q}}
\vec{\sigma }c _{{\bf k}}
$ is the electronic spin density and $\vec{M}_{{\bf q}}$ 
represents the amplitude of the spin-density wave at
wavevector $\bf q$. 
In the paramagnetic Fermi liquid, the spin 
fluctuations have a finite correlation length and correlation time:  
it is the virtual emission  of these soft fluctuations via the process
\begin{equation}\label{}
e^{-} \rightleftharpoons e^{-}+ \hbox{spin fluctuation}
\end{equation}
that then gives rise to mass renormalization. 
Ultimately, once the magnetic order develops, 
electron Bragg scattering off the spin density wave
causes the Fermi surface to ``fold'' along lines in
momentum space.  In this picture, the electrons which form the
Fermi surface on the paramagnetic and the antiferromagnetic side
of the quantum critical point are closely related to one-another. 

The alternative ``strong coupling'' response to these questions 
treats the heavy fermion metals as a
Kondo lattice of local moments\cite{doniach}.  From this perspective,
heavy electrons are {\sl composite} bound-states formed 
between local moments and high energy conduction
electrons. Here, 
the underlying spinorial 
character of the magnetic fluctuations plays a central role
in the formation of heavy quasiparticles. For instance,
the Fermi surface volume, or Luttinger sum rule\cite{luttinger}  
in the paramagnetic phase  ``counts'' both the
number of conduction electrons {\sl and} the number of heavy-electron
bound-states, given by 
the number of spins:
\begin{equation}\label{}
2\frac{{\cal V}_{FS}}{(2\pi)^{3}}= n_{e} + n_{spins}
\end{equation}
where ${\cal V}_{FS}$ is the volume of the Fermi surface, $n_{e}$ is the number of conduction electrons per unit cell
and $n_{spins} $ is the number of spins per unit cell.\cite{martin,oshikawa}
This scenario departs fundamentally from the 
spin-density wave scenario if we suppose that 
at the critical
point, the bound-states which characterize the Kondo lattice
disintegrate.  \begin{figure}[here]
\[
\figx{0.8\linewidth}{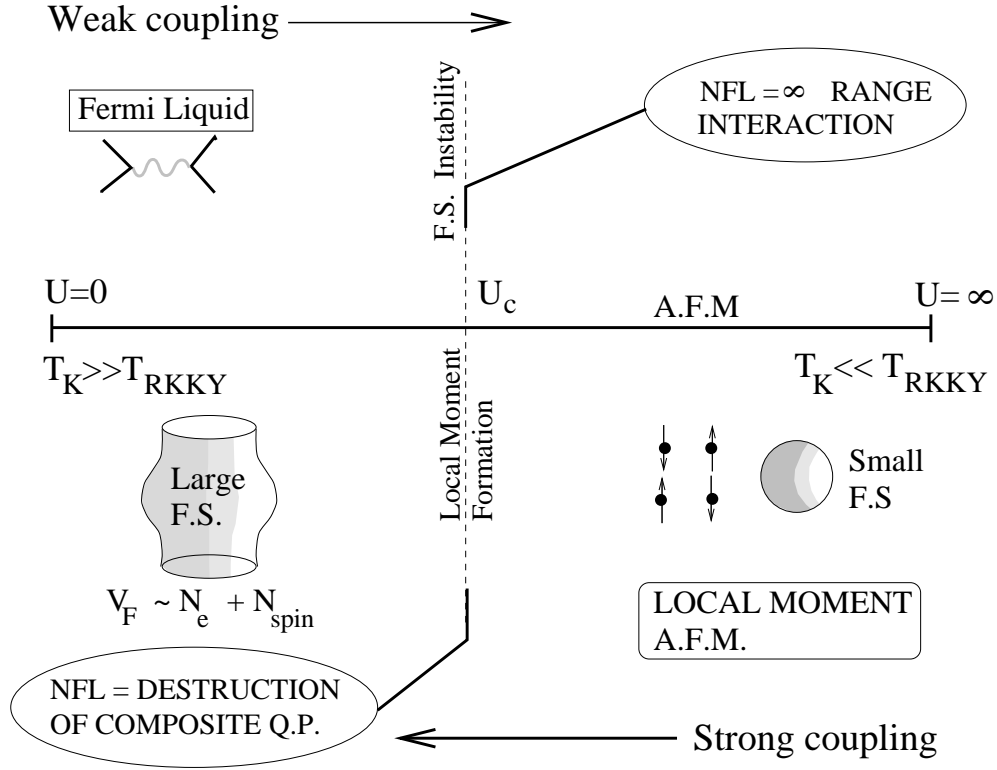}
\]\vspace{-4mm}
\caption{\label{fig2}Contrasting the weak, and strong-coupling picture
of an antiferromagnetic QCP.  }\end{figure} 
%\newpage
%\noindent
\noindent 
In this case, the Fermi liquid in the paramagnetic
and magnetic phases involve  different 
electron quasiparticles. The antiferromagnet involves a fluid
of conduction electrons immersed in a lattice of ordered moments.
By contrast, the heavy fermion paramagnet involves quasiparticles
built from a bound state between conduction electrons and the local
moments: the magnetic degrees of freedom are confined and manifest 
themselves as new spinorial  degrees of freedom.
The Fermi surface  reconfigures at the QCP to accomodate the changing 
character and density  of the quasiparticles. 
From this point of view, it is natural to suppose that 
the spinorial character of the magnetic  degrees of freedom seen in the
heavy fermion phase will {\sl also }manifest itself
in the decay modes of the heavy quasiparticles. 
For example, 
the heavy quasiparticles could decay into a neutral
``spinon'' and a spinless, charge $e$ fermion, schematically
$e^{-}_{\sigma }\rightleftharpoons s_{\sigma } + \chi ^{-}$,
corresponding to 
\begin{align*}
L_{F-M}^{(2)} = g \sum _{{\bf k}, {\bf q}}[ s\dg _{{\bf k}-{\bf
q}\sigma}\chi _{\bf q} \dg c_{{\bf
k}\sigma }  +\hbox{H.c}],
\end{align*}
where $s\dg _{{\bf q}}$  is a neutral spin-$1/2$  boson and $\chi $ 
a spinless charge $e$ fermion,  reminiscent of a hole in 
a Nagaoka ferromagnet.\cite{thouless,nagaoka}
Indeed, if the magnetism enters as  a spinorial  field, its
coupling
to the electron field can only occur via  an inner product over the spin
indices
as shown in $L_{F-M}^{(2)}$.  The collective magnetic correlations 
of the spinor
field oblige us to cast it as a boson, and likewise, 
statistics forces us to introduce the 
additional spinless fermion into the coupling.  In other words, the
idea that magnetism enters into the decay modes as a spinorial field
constrains the coupling Lagrangian to the above form.

In this second  picture, the scale
$T_{F}^{*}(P)$ is the threshold energy above which the composite
particles decay into their constituent particles
and magnetism develops via the 
condensation of the spinon field. This in turn,  transforms the Fermi surface by opening up a resonant channel between
the heavy electrons and the spinless fermions. 

Let us now examine these alternatives
in greater detail (see figure 2). 
\fight=0.6\textwidth
\fg{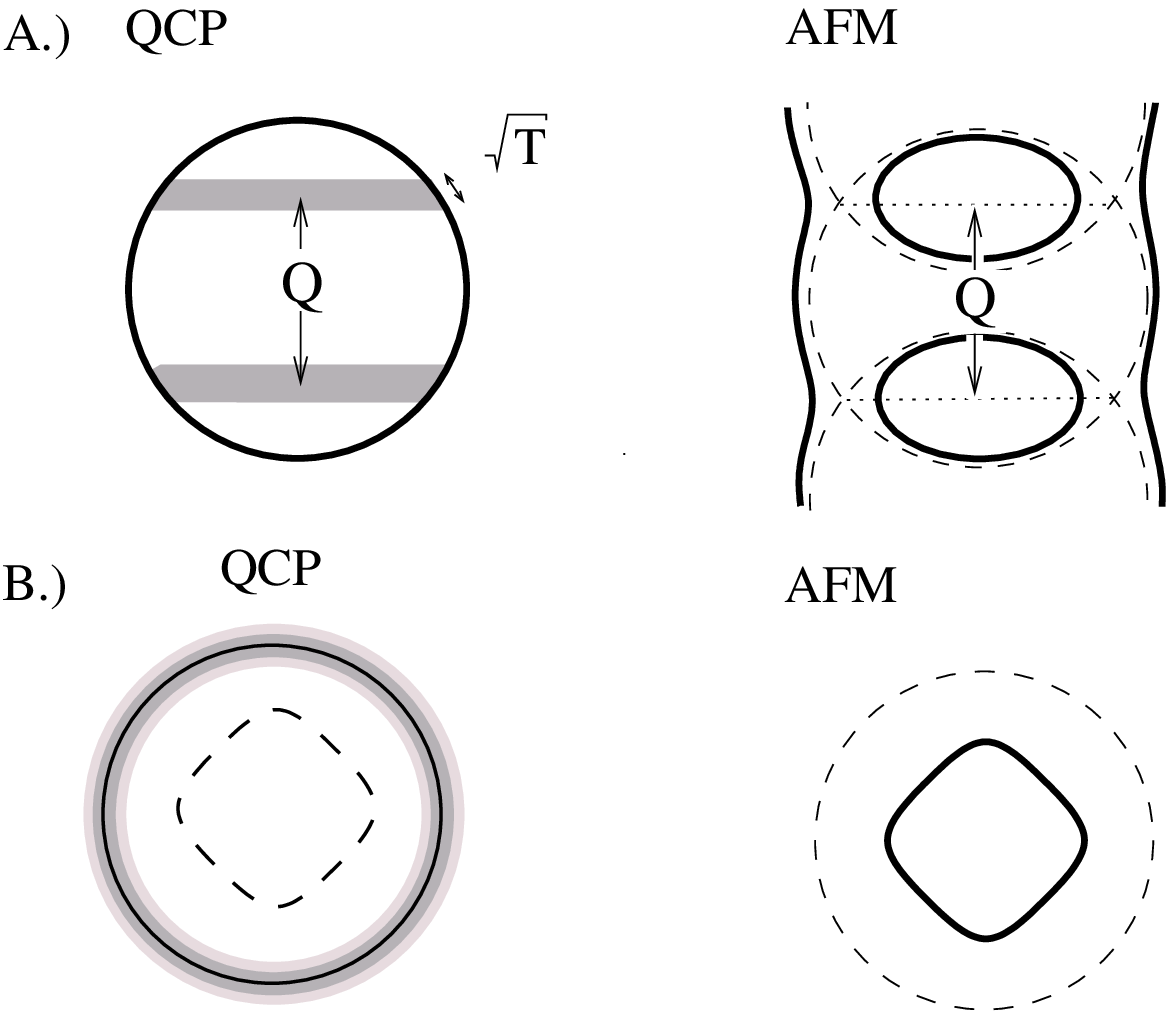}{fig3}{Competing scenarios for the
antiferromagnetic QCP in heavy fermion materials (a) spin density wave
scenario, where the Fermi surface ``folds'' 
along lines separated by
the
magnetic 
$Q$ vector, pinching off into two separate Fermi surface sheets; 
(b) sudden reconfiguration of the fermi-surface
accompanies
break-down of composite heavy fermions.  The fermi surface in the
antiferromagnetic phase only incorporates the conduction electrons.
}

In the spin-density wave picture, 
the internal structure of the quasiparticles is unimportant:
the physics is entirely described by the interaction between the
Fermi surface and critical antiferromagnetic spin fluctuations
(Fig 3(a)) Bragg reflections off these critical fluctuations 
strongly couple the lines of excitations on  
the Fermi surface that are separated by the 
critical wavevector ${\bf Q}$. Along the hot lines, the electron energies
are given by
\begin{equation}\label{}
\epsilon_{\bf k}=\epsilon_{{\bf k}-{\bf Q}}=0.
\end{equation}
Beyond the critical point, the Fermi surface 
folds  along these ``hot lines'', pinching into two separate 
and smaller Fermi surfaces, as
shown in (Fig 3 (a)). 
At the quantum
critical point, quasiparticles along the ``hot lines'' 
are critically scattered with divergent scattering
rate and effective mass.

The quantum critical behavior predicted by this model has been extensively
studied,\cite{hertz,millis} 
and the effective ``Hertz-Millis'' action describing the critical fluctuations takes the form
\begin{equation}\label{}
L_{M} = \sum_{i\omega_{n}}\int \frac{d^{3}q}{(2\pi)^{3}}\vert M
(q,\omega_{n})\vert ^{2}
\chi _{o}^{-1} ({\bf q},\omega_{n})
+ U\int d^{3}r d\tau M ({\bf r},\tau )^{4}
\end{equation}
where 
\begin{equation}\label{}
\chi _{o}^{-1} ({\bf q},\omega_{n})=
\left [ 
({\bf q}-{\bf Q} )^{2} + \xi^{-2} + \vert
\omega_{n}\vert/\Gamma_{\bf Q}
\right]  \chi _{0}^{-1}
\end{equation}
is the inverse dynamical spin susceptibility of the magnetic fluctuations.
The linear damping rate of the magnetic fluctuations is derived from
the density of particle-hole excitations in the Fermi sea.  $\xi^{-1}\sim
(P-P_{c})^{\frac{1}{2}}$
is the inverse spin correlation length, whilst $\tau ^{-1}= \Gamma
_{{\bf Q}}\xi^{-2}$ is the inverse spin correlation time. 
An important feature 
of this ``$\phi^4$'' Lagrangian is that the momentum dependence enters
with twice the power of the frequency dependence and
$
\tau\sim\xi^z,
$
where $z=2$ (the dynamical critical exponent of this theory), so that
the time dimension counts as {\sl two } space dimensions, and 
the effective dimensionality
of the theory is 
\begin{equation}\label{}
D= d + z = d+2
\end{equation}
Assuming $d=3$ in heavy fermion systems then  $D=5$ exceeds the
upper critical dimension $D_{c}=4$ for a $\phi^4$ theory. 
This has three immediate consequences\begin{itemize}

\item 
the interactions amongst the critical modes are  ``irrelevant'', 
scaling to zero at large scales, so that 
the long-wavelength
antiferromagnetic modes are non-interacting 
Gaussian modes, or over-damped ``phonons''.  The absence of
non-linearities
in the interactions means that singularities in the magnetic
response will remain confined to the region around the Bragg point,
and will not manifest themselves in the uniform susceptibility.

\item 
the temporal correlation time entering into 
the 
magnetic response functions will involve $E/T^{3/2}$, rather than
$E/T$ scaling \cite{sachdev} (see earlier discussion).

\item the correlation functions
and thermodynamics will not exhibit anomalous scaling exponents.

\end{itemize}
The first point is difficult to reconcile with the observation of
a singular temperature dependence in the  uniform magnetic susceptibility.
The second and third points are 
incompatible with the divergence of the
specific heat and the observation of $E/T$ scaling with anomalous exponents\cite{schroeder}.
To gain more insight into the physics that lies behind 
these  difficulties, let us examine the nature
of the spin fluctuations predicted in this picture. 
The Gaussian 
critical spin fluctuations predicted by the quantum spin density wave
picture 
mediate an effective interaction
given by 
\begin{equation}\label{}
V_{eff} ({\bf q},\omega )= g^{2}\frac{\chi _{0}}{({\bf q}-{\bf
Q})^{2}-\frac{i\omega }{\Gamma_{\bf Q}}}
\end{equation}
In real-space, this corresponds to a ``modulated '', but unscreened 
Coulomb potential
\begin{equation}\label{}
V_{eff} (r,\omega =0)\propto \frac{1}{r}e^{i {\bf Q}\cdot {\bf r}}
\end{equation}
The rapid modulation in the above
potential produces Bragg scattering. Unlike 
a critical ferromagnet, where  the singular
scattering potential  affects all points
of the Fermi surface, here the modulated  potential
only couples electron quasiparticles along 
``hot lines'' lines on the Fermi surface
that are separated by the wave-vector $\bf Q$.  

In practice, 
thermal spin-fluctuations of frequency $\omega \sim k_{B}T$ are
excited, so that electrons within a  strip of momentum
width 
\begin{equation}\label{}
\Delta k \sim \sqrt{\frac{k_{B}T}{\Gamma_{{\bf Q}}}}
\end{equation}
around the hot lines are strongly scattered by the critical
fluctuations. The spin-fluctuation
self-energy produced by these critical fluctuations is denoted by the
Feynman diagram
\fight=0.4\textwidth
\bxwidth=0.4\textwidth
given by 
\begin{equation}\label{}
\Sigma ({\bf k},\omega )= -T \sum_{{\bf q},\nu }g^{2} \chi_{o} ({\bf
q},\nu ) G ({\bf k}-{\bf q},\omega -\nu )
\end{equation}
%\fg{fig4.eps}{fig4}{Exchange self-energy}
\begin{center}\frm{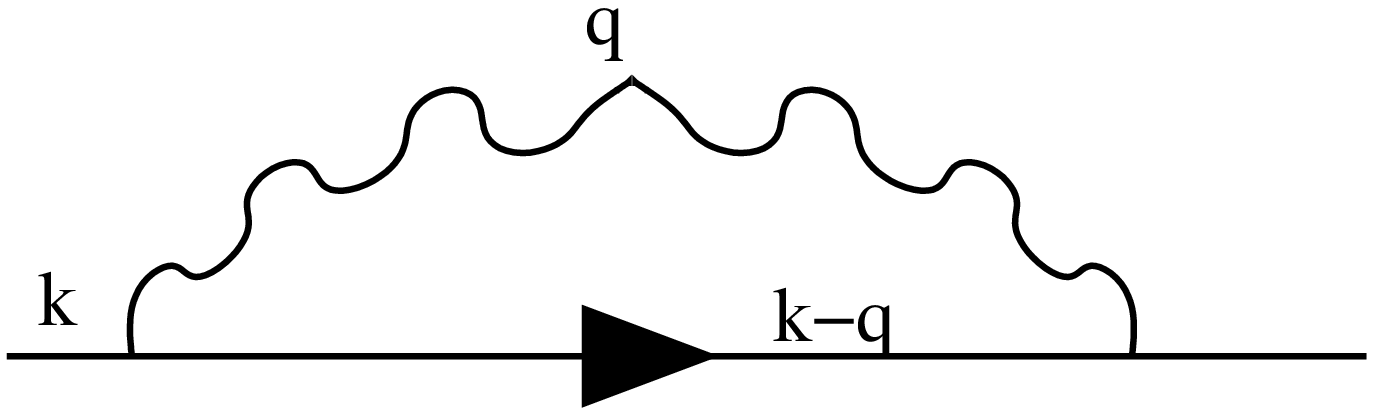}
\end{center} 
\noindent where $g$ is the strength of the coupling between spin fluctuations and
conduction electrons and $G ({\bf k},\omega)=[\omega -\epsilon _{{\bf k}}]^{-1}$ is the Green function of
the conduction electrons.  Along the hot lines, this produces a 
Marginal self-energy of the form $\Sigma ({\bf k}_{hot},\omega)\sim
\omega {\ln} [\hbox{max} (\omega,T)/\Gamma] $, giving rise to a mass renormalization
with a weak logarithmic divergence
\begin{equation}\label{}
\frac{m^{*}}{m}\sim 1 - \frac{\partial \Sigma }{\partial \omega }\sim
\ln (\Gamma /T)
\end{equation}
This weak logarithmic divergence only extends along the narrow
``hot band'' of width $\Delta k\sim \sqrt{T}$: elsewhere on the Fermi
surface the electrons are essentially unaffected by the critical fluctuations.
For this reason the critical spin fluctuations should only produce a weak
singularity in the specific heat, if the quantum critical behavior
is described by a spin density wave instability. 
The  residual
contribution to the specific heat produced by this narrow band of
non-Fermi liquid behavior is expected to depend on $\sqrt{T}$. This
can been seen by noting that 
the singular contribution to the Free energy  at the quantum
critical point is due to the gaussian fluctuations of the
critical antiferromagnetic modes, given by
\begin{eqnarray}\label{lab2}
F_{sing} &=& T \sum_{i\omega_{n}}\int \frac{d^{3}q}{(2\pi)^{3}}
\log [ \chi _{o}^{-1} ({\bf q},\omega_{n})]\cr
&=& \int \frac{d^{3}q}{(2\pi)^{3}} \int \frac{d\nu}{\pi}
\left[ n(\nu)+ { \frac{1}{2}}\right]
\tan
^{-1}\left[\frac{\nu /\Gamma_{\bf Q}}{[({\bf q}-{\bf Q} )^{2}]} \right]
\end{eqnarray}
By rescaling $\nu = u T$ and $({\bf q}-{\bf Q})
={\bf x} \left(\frac{{T}
}{
\Gamma_{\bf Q}}\right)^{\frac{1}{2}} $, we see that 
\begin{eqnarray}\label{lab3}
F_{sing} &=& \frac{T^{\frac{5}{2}}}{\Gamma
_{\bf Q}^{\frac{3}{2}}}\int \frac{d^{3}x}{(2\pi)^{3}} \int \frac{du }{2 \pi}
\left[ \coth (u/2) \right]
\tan^{-1}\left[\frac{u}{{\bf x} ^{2}  } \right]
\end{eqnarray}
scales as $T^{{5/2}}$, giving rise
to a contribution to the linear specific heat coefficient 
$\partial^{2}F/\partial T^{2}\sim \sqrt{T}$. Thus the specific
heat at the QCP of a critical three dimensional spin density contains
a singular $\sqrt{T}$ component, but it is
\underline{not divergent}.  

Some aspects of the above picture are modified by
disorder. Rosch\cite{rosch2} has argued
that a disorder can significantly modify  the
transport properties of the spin density wave picture. Thus in a strictly
clean system, the resistivity should be short-circuited by the electrons
far from the hot lines, giving rise to a $T^{2}$, rather than a
quasi-linear resistivity. High temperatures plus disorder give rise to a resistivity 
dominated by the average  scattering rate, given approximately by
a product of the width of the hot line and the linear scattering rate
on the hot line ($\sqrt{T}\times T= T^{1.5}$).  Rosch has pointed out that
the cross-over between these two limits is extremely broad, and gives
a quasi-linear resistivity. However, this still does not account for
the divergence of the specific heat and the $E/T$ scaling.

One proposed resolution to this
paradox is to suppose that the spin fluctuations in the
heavy fermion quantum critical systems form a two, rather
than
three dimensional spin fluid. 
If,  for example, due to the effects
of frustration, over the observed temperature range, the spin
fluctuations were confined to decoupled planes in real space, then the
critical spin physics would be two-dimensional and the effective
dimensionality of the quantum critical point would be $2+z=4$, putting
the fluctuations at their critical dimensionality.  This explanation
was first advanced by Rosch et al to account for the logarithmic
dependence of the specific heat coefficient in $CeCu_{6-x}Au_{x}$,
(x=0.1)\cite[]{rosch}.  Inelastic neutron scattering experiments
on this material  do provide circumstantial support for reduced
dimensionality, showing that the 
critical scattering appears to be concentrated
along linear, rather than at point-like regions in reciprocal space\cite[]{schroeder,rosch}. 
%%%MOVED TO LATER. SEE RATIONALE BELOW
%This however, is not sufficient to explain the
%anomalous exponents in the neutron
%data.  Si et al~\cite{si} have recently proposed that 
%quasi-two dimensional
%spin fluctuations interact with the Kondo effect to produce a
%``local quantum  critical point '' 
%which give rise to localized spin fluctuations with critical
%correlations in time that exhibit $E/T$ scaling.
In an independent
discussion, Mathur et al~\cite{mathur} have suggested that the spin fluctuations
in  quantum critical $CePd_{2}Si_{2}$ and $CeGe_{2 }Ni_{2}$ might
be driven to be two dimensional by frustration. 

%%%MOVED HERE
This two-dimensional SDW picture, however, cannot explain
the anomalous exponents -- both at and far away from the ordering
wavevector ${\bf Q}$ -- in the neutron and magnetization data.
Together with the $E/T$ scaling, the 
experiments instead suggest a fundamentally new {\it interacting} fixed 
point. There are two approaches in the search for such a 
new universality class. 
Si et al~\cite{si} have recently proposed that 
quasi-two dimensional
spin fluctuations interact with the Kondo effect to produce a
``local quantum  critical point '' which gives
rise to localized spin fluctuations with critical
correlations in time that exhibit $E/T$ scaling.
This picture raises many questions. Do the quasi-2D
fluctuations, seen in $CeCu_{6-x}Au_{x}$, also occur
in the other heavy fermion systems with a divergent specific
heat coefficient ? %%%THE NEXT QUESTION IS A NON-ISSUE ONCE THE PREVIOUS ONE IS ANSWERED.
%%%SURELY CUPRATES SERVE AS A PERFECT EXAMPLE OF 2D FLUCTUATIONS BUT 3D ORDER
%
%%- No!  The cuprates are far more 2D electronically than heavy
%% fermion systems- yet these systems DO show a cross-over to 3D
%% behavior- indeed- when it occurs, three d order develops.
%% In the context of the heavy fermions, we expect a cross-over from
%% 2D to 3D quantum critical behavior at some point- why is it not observed?
More microscopically,
why should quantum critical heavy fermion systems
have a tendency towards quasi-two dimensional spin fluids,
when the transport is highly three dimensional?
%However, the reduced dimensionality picture 
%raises many
%questions. 
%In particular- why should quantum critical heavy fermion systems
%have a tendency to form quasi-two dimensional spin fluids, when the transport
%is highly three dimensional.

%The alternative and competing 
On the other hand, Coleman, Pepin and Tsvelik\cite{susy}, advance
the view
%is 
that the anomalous quantum critical behavior seen in these systems
is a feature of a truly three dimensional spin system, but one governed 
by a new class of quantum critical behavior with upper critical
dimension larger than three.  But this too raises many questions-
in particular- can a field theory of the form (\ref{lag}) be found
in which the upper critical spatial dimension is greater than three, and what
experimental  signatures would this lead to?

The two issues - whether the quantum criticality in heavy fermions
is quasi-two dimensional
%%%ONCE AGAIN TO ME THERE ARE TWO QUESTIONS INSTEAD OF ONE
%or whether - OK- note the plural 
or purely three dimensional,
and whether or not
it represents a 
fundamentally
new class of
quantum critical behavior, lie at the heart
of the current debate about these systems.  Experimentally,
it remains a task of high priority to examine the momentum
distribution of the critical scattering, 
%%% THE ADDITION REFLECTS ON THE TWO QUESTIONS
and the extent to which $E/T$ scaling prevails,
in stoichiometric quantum critical heavy fermion systems.
More detailed information on 
manifestly three-dimensional systems, such as cubic quantum-critical
$CeIn_{3}$ would also be useful in this respect.

\subsubsection{Does the Hall Constant Jump at a QCP?}

We should like to end this discussion by asking 
what other independent experimental signatures, over and beyond the
divergence of the linear specific heat  and neutron scattering, 
can be used to 
test the mechanism by which the Fermi surface transforms between the
paramagnet and the antiferromagnet?  Here, one of the most promising,
but surprisingly, untested probes is the Hall constant. 
If the same quasiparticles are involved in the
conduction process on both the paramagnetic and the antiferromagnetic
side of the quantum critical point, then
we expect the Hall constant to vary continuously through the
quantum phase transition. If by contrast, the entire character of the
Fermi surface changes, for instance, via the creation or 
destruction of fermionic
resonances at the Fermi energy, then we expect that the Hall constant
will encounter a discontinuity at the quantum critical point. In the most
extreme example, it may even change sign.

To see the logic in this discussion more clearly, consider first the
case of a spin density wave.  In a Boltzmann transport approach, 
the Hall conductivity of 
the Fermi surface is  \cite{ong}
\begin{equation}\label{}
\sigma _{xy}\propto \int dk_{z}\int \left(\frac{{\bf v}
\times d{\bf v}}{\Gamma_{tr}^{2} ({\bf k})} \right)
\end{equation}
where ${\bf v}= \nabla _{{\bf k}}E_{\bf k}$ is the group velocity
on the Fermi surface. The above integral corresponds to the average
area swept out by the group-velocity vector around the Fermi surface. 
When the spin density wave develops, Bragg reflections cause 
the Fermi surface to fold 
and pinch into two separate  sheets, as illustrated in
%%%I THINK THE FIGURE NUMBER IS WRONG
Fig. 3. 
The group-velocity on the two separate Fermi surfaces that develop is determined by
the renormalized energy 
\begin{equation}\label{}
E^{\pm }_{\bf k}= \frac{1}{2} (\epsilon_{\bf k}+ \epsilon_{\bf k+Q })\pm
\sqrt{
\bigl(\frac{\epsilon_{\bf k}- \epsilon_{\bf k+Q }}{2}\bigr)^{2}
+ g^{2}M^{2} _{\bf Q}}
\end{equation}
where $\pm$  refers to the dispersion on the two separate sheets, 
$M_{\bf Q}$ 
is the staggered magnetization at wavevector $\bf Q$ and
$\epsilon_{{\bf k}}$ is the quasiparticle dispersion in the heavy
fermi liquid.
Notice that away from the hot lines, the change in the group velocity
induced by the magnetic order is second order in the staggered
magnetization and that furthermore, the Fermi surface volume 
is conserved through the transition
\begin{eqnarray}\label{sdw}
\Delta v_{\rm FS} &=&0\cr 
\nabla_{\bf k} E^{\pm }_{\bf k}& =& \nabla_{{\bf k}} 
\epsilon_{\bf k} + O (M_{\bf Q}^{2})
\end{eqnarray}
Away from the hot lines we also expect the scattering rate 
to change no faster than the square of the magnetization. 
Near the quantum critical point, the Fermi surface becomes
increasingly sharp in the vicinity of the hot lines which 
develop into ``corners'' of 
the Fermi surface.
%%%I THINK THE FIGURE NUMBER IS WRONG HERE AS WELL
(Fig. 4.) Approaching  the quantum critical point
from the magnetic side, the Hall conductivity can be divided into two
parts
\begin{equation}\label{}
\sigma _{xy}\sim \left\{\int_{sheets} dk_{z}+ \int_{corners} dk_{z}\right\}
\int \left(\frac{{\bf v}
\times d{\bf v}}{\Gamma_{tr}^{2} ({\bf k})} \right)
\end{equation}
\fight=0.8\textwidth
\fg{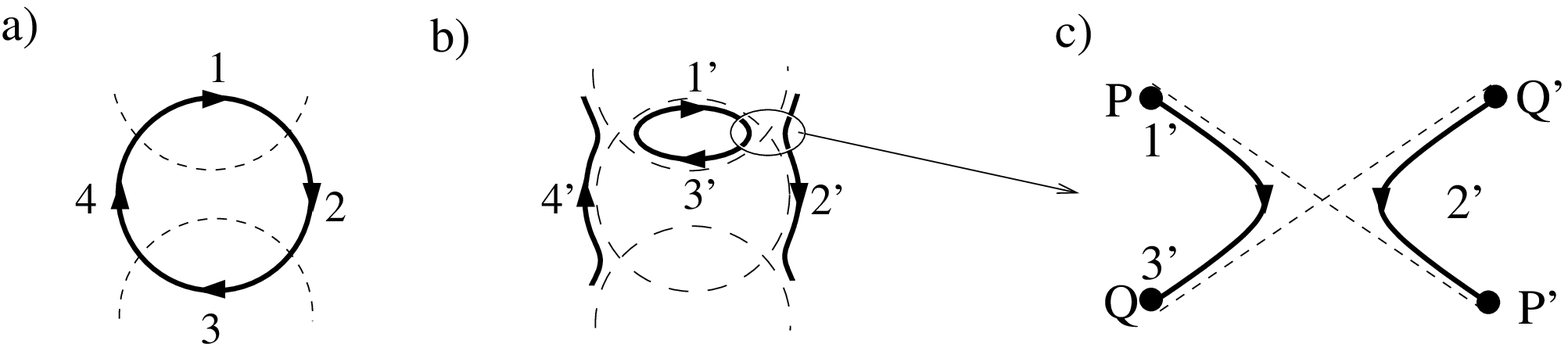}{fig5}{
Fermi surface in (a) paramagnetic phase  and 
(b) spin density wave phase, showing how
regions 1-4 have folded and pinched to form two new Fermi surfaces.
(c) Expanded detail of the ``corners'' of the Fermi
surface
in the vicinity of the saddle point along the hot line. Points $P$ and
$P'$,  Q and Q' correspond to quasiparticles with the same group
velocity and scattering rate, but equal and opposite effective masses.
Thus the integrals $\int_{P}^{Q} d{\bf l} \times {\bf l}
= - \int
_{Q'}^{P'}d{\bf l} \times {\bf l}
$ and the total contribution to the Hall conductivity from the 
corners of the Fermi surfaces identically cancel. 
}
\noindent The contribution to this integral from the ``sheets'' of the Fermi
surface converge to that of the paramagnetic Fermi liquid. The 
corners of the Fermi surface correspond to saddle points in the
electron dispersion, and close to the quantum critical point the
corner regions on the two daughter Fermi surfaces are exact mirror images
of one-another with identical scattering rates but equal and opposite group velocities and mass
tensors:  for this reason
the contributions to the Hall conductivity 
from the saddle point region of one Fermi surface identically cancel
the corresponding contribution from the corners of the second. (Fig 4
(c))
In the remaining portions of the Fermi surface, the change in the
quasiparticle energy and group velocity is proportional
to $M_{\bf Q}$. 
Since the change in dispersion of the electrons away from the hot
lines is dependent on the square of the magnetization, we expect
the change in the 
Hall constant to grow quadratically with the staggered
magnetization 
a quantum critical point is driven by diffraction off a spin density
wave. 
\begin{equation}\label{}
\Delta R_{H}\propto M_{\bf Q}^{2}, \qquad \qquad (S.D.W.)
\end{equation}
which in the simplest models, with $M_{\bf Q}\propto \vert
P-P_{c}\vert ^{\frac{1}{2}}$ would imply a jump  in 
$dR_{H}/dP$, but no jump in $R_{H}$.
The continuity  of the Hall constant in this scenario is a 
direct consequence  of the continuity in the quasiparticle character,
through the transition.

By contrast, if a  break-down of composite heavy fermions develops 
at the quantum critical point, then 
we expect  the more abrupt changes in the Fermi surface
to manifest themselves in the Hall constant. 
In the simplest possible models of heavy electron behavior,
the quasiparticles in the Kondo lattice paramagnet are holes
with a Hall constant that is opposite in sign to the conduction
electrons from which they are formed. If the change in character
from conduction electron to heavy electron were to take place at the
quantum critical point, 
the Hall constant might not just jump- it could
even change sign at the quantum critical point. To emphasize this
more fully, consider the following toy model 
in which  the spinorial character of the magnetism appears in
$L_{FM}$, 
\begin{align*}
H= \sum_{{\bf k}\si}\epsilon_{\bf k}c\dg _{{\bf k}\si}c _{{\bf k}\si}
+g \sum _{{\bf k}, {\bf q}}[ s\dg _{{\bf k}-{\bf
q}\sigma}\chi _{\bf q} \dg c_{{\bf
k}\sigma }  +\hbox{H.c}] + \lambda \sum _{\bf k} \chi\dg _{\bf k} \chi_{\bf k}
+L_{M}[s]\ ,
\end{align*}
where $\lambda $ is the chemical potential of the spinless fermions.
Antiferromagnetism results from a condensation of the spinorial field, 
$\la s_{{\bf
q}\si}\ra
=\sqrt{2M_{\bf Q}}z_{\si}\delta _{{\bf q}-\sigma {\bf Q}/2}$, so that the effective 
effective Hamiltonian 
takes the form
\begin{equation}\label{}
H= \sum_{{\bf k}\si}\epsilon_{\bf k}c\dg _{{\bf k}\si}c _{{\bf k}\si}
+g \sqrt{M_{\bf Q}}\sum _{{\bf k}\sigma}
[ \chi _{\bf k- \sigma Q/2} \dg c_{{\bf
k}\sigma }  +\hbox{H.c}] + \lambda \sum _{\bf k} \chi\dg _{\bf k} \chi_{\bf k}
+L_{M}[s]\ .
\end{equation}
Once the magnetization develops, the
the hybridization with the $\chi $ fermions has two effects. First, it
changes the dispersion of the up and down spin quasiparticles 
according to 
\begin{equation}\label{}
\omega^{(\pm)}_{{\bf k}} = \frac{\epsilon_{{\bf k}}+\lambda}{2}\pm
\sqrt{[( \epsilon_{{\bf k}}-\lambda)/2]^{2}+ g ^{2 }M_{{\bf Q}}}.
\end{equation}
Notice that the change in the dispersion is now first order in the
magnetization. 
Second, the staggered magnetization induced by the condensation of
the spin-bosons causes Bragg diffraction which mixes up
and down  bands to producing fermions with dispersion
$E_{{\bf k}}$ determined from the roots of the equation
\begin{equation}\label{}
\prod _{\pm}(E - \omega^{(\pm)}_{{\bf k}}
)
(E - \omega^{(\pm)}_{{\bf k}+{\bf Q}})
= ( g^2 M_{{\bf Q}})^2
\end{equation}
There are two main consequences of the disintegration of the heavy
fermions-  first,
the Fermi surface volume now has to change to accomodate
the resonances formed by the disintegration of the heavy electrons
and second, the change in the dispersion is {\sl first order}
in the  staggered magnetization:
\begin{eqnarray}\label{lab4}
\delta v_{{\rm FS }}&=& n_{\chi }/2\cr
\nabla_{\bf k} E_{{\bf k}}&=& \nabla \epsilon_{{\bf k}} + O (M_{\bf Q})
\end{eqnarray}
Suppose the spinless  fermions 
are in equilibrium
with the heavy quasiparticles at the Fermi surface, so that
the chemical potential $\lambda=0$ 
at the quantum critical point, in this case 
$n_{\chi }$
starts out with a finite value, and the Fermi surface volume
jumps at the transition. 
If by contrast, 
$\lambda \ne 0$ at the transition, then $n_{\chi }$ would start out
from zero, but the change in the dispersion would still be first order
in the staggered magnetization.  In the former scenario, the Hall constant
jumps at the transition, whereas in the latter it evolves in direct
proportion to the change in the staggered magnetization. 
Summarizing:
\begin{eqnarray}\label{lab5}
\Delta R_{H}\propto  \left\{\begin{array}{cr
}
O (1)& \qquad (\lambda=0)\cr
M_{\bf Q}&\qquad (\lambda \ne 0)
\end{array} \right.\qquad (\hbox{composite heavy fermions})
\end{eqnarray}
The main point (Table 2.  ) is that, a large discontinuity in the gradient or jump
in the Hall constant is expected in a picture where 
the 	quantum critical point
involves a break-down of composite heavy fermions, whereas 
a spin density wave picture predicts a continuous change in the
Hall constant with a finite jump in $dR_{H}/dP$ at the quantum
critical point.

\subsubsection{Conclusion}

To conclude, we have discussed the challenge posed by the apparant
divergence in quasiparticle masses that appears to develop at a heavy fermion
quantum critical point.  We have emphasized in some length how this poses
severe difficulties for a model based on the development of a spin
density wave instability.  
%%%ONCE AGAIN, TO ME THERE ARE TWO QUESTIONS INSTEAD OF ONE. AND IN LINE WITH
%%%THE OVERALL TONE OF THE PAPER, I'VE RE-PHRASED THE FIRST SENTENCE FROM
%%%A STATEMENT TO A QUESTION.
%% - OK- PIERS
%Many of the observed properties suggest
%that the underlying quantum critical physics involve a more complete
%break-down in quasiparticle physics than that expected from a quantum
%spin density wave instability. - OK- PIERS
%!This has set the ground for a lively debate
%!between two different theoretical scenarios- one- in which 
%1the critical spin fluctuations in these materials are asserted to be quasi-two
%!dimensional,  the other, in which the unusual non-Fermi liquid physics is
%!taken to signal a fundamentally 
%!new class of three dimensional quantum critical behavior.
%% Qimiao- the word ``asserted'' has to remain here, because any other
%% word would suggest that quasi-two dimensional behavior is an
%% accepted feature of heavy fermion quantum critical behavior-
%% I don't even accept this idea for Cerium Copper-6 gold! So- I 
%% do insist on ``asserted'' staying where it is. - PIERS
\newpage
\bxwidth=0.8\textwidth
\begin{center}
\frm{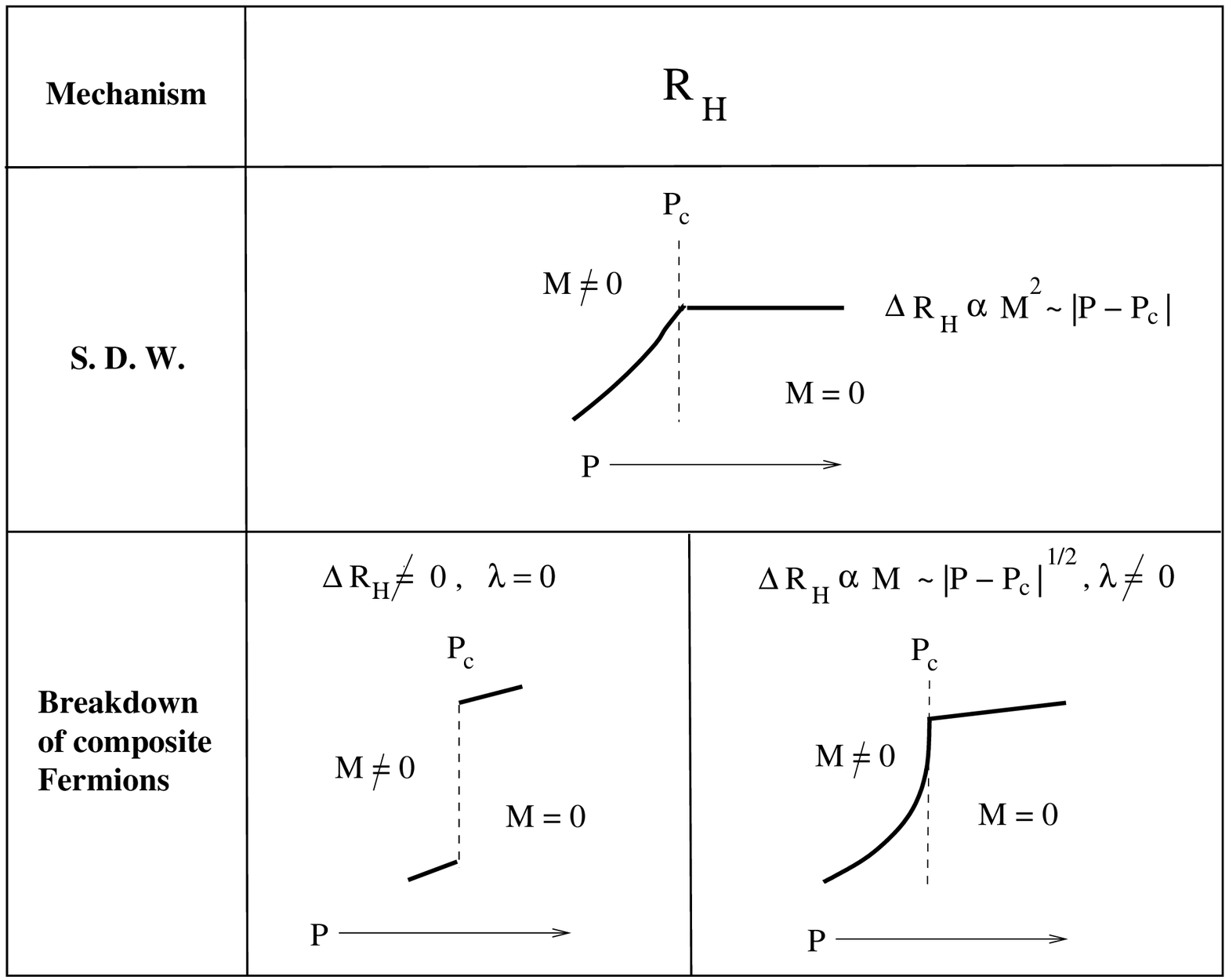}
\end{center}
%\fg{fig6.eps}{fig6}{
\vskip 0.3truein \noindent {\bf Table 2.} 
Variation of the Hall constant expected in a spin
density wave scenario, and a scenario where the composite heavy
fermions
disintegrate at the quantum critical point. For the purposes of
illustration, it is assumed that the magnetization grows as the
square-root of the control parameter, 
$M\propto \sqrt{\vert P-P_{c}\vert }$.
\vskip 0.2 truein
This has set the ground for a lively debate on two different fronts.
First and foremost, do the experiments require a more 
complete break-down in quasiparticle physics than that expected from
a quantum spin density wave instability?
Second,  are the quantum critical fluctuations quasi-two dimensional
in character, or is the quantum critical behavior signal 
a new class of three dimensional quantum criticality?
%does the quasi-two dimensional nature of the spin fluctuations
%play a crucial role? 
It is hoped that these
discussions will stimulate further experimental work .  In particular,
further experimental 
clarification would be immensely useful in two  separate  respects:

\begin{itemize}

\item Further evidence for the divergence of the linear specific heat
and the collapse of the Fermi temperature 
near a heavy fermion quantum critical point is needed. 
More direct specific heat capacity measurements and further measurements of the quadratic coefficient of the
resistivity in the
approach to the quantum critical point would help to elucidate
this issue. 

\item Careful examination of the evolution of the Hall constant 
at a heavy fermion quantum critical point will provide the means
to check directly whether the Fermi surface ``folds and pinches''
under the influence  of Bragg diffraction off a spin density wave, 
with no discontinuity in the Hall constant, or
whether it undergoes a fundamental transformation due the introduction
of new fermionic resonances into the Fermi sea, producing a
jump in the Hall constant. 

\end{itemize}

\noindent {\bf Acknowledgments}

%%%QS ACKNOWLEDGE INSERTED
This work was supported in part by the National Science Foundation
under grants DMR 9983156 (PC), DMR 0090071 (QS)and PHY
99-07947 (PC, CP and QS). The work of Q.S. was also supported by 
a grant from TCSUH. 
The work of R. R.  was supported by the MacArthur 
Chair endowed by the John D. and Catherine T. MacArthur
Foundation at the University of Illinois.
PC would like to thank G. Aeppli for many intensive discussions
on aspects of this work, particularly for sharing with him
experimental measurements on the Hall constant at quantum critical
point, prior to publication. 
PC is also grateful for discussions on matters
relating to this work with   A. Chubukov and G. Lonzarich.
C.P acknoledges discussions with C. De Dominicis, J. Zinn-Justin and
J. M. Luck. QS would like to thank K. Ingersent, S. Rabello, and L. Smith
for an on-going collaboration on the subject.
PC, CP  and QS would like to thank the Newton Institute, Cambridge, UK
and the Institute for Theoretical Physics
Santa Barbara, where part of this work was carried out.

\maketitle

\end{document}